\documentclass[usenatbib]{mn2e} 
\usepackage{psfig}
\usepackage{lscape}
\usepackage{rotating}
\usepackage{psfig}
\usepackage{amsmath}
\usepackage{amssymb}
\title[Impact of the warm outflow in PKS 1345+12]{The impact of the warm
  outflow in the   young (GPS) radio source \& 
  ULIRG PKS 1345+12 (4C 12.50)}
\author[J. Holt et al.]{J. Holt$^{1}$\thanks{E-mail:
jholt@strw.leidenuniv.nl}, C. N. Tadhunter$^{2}$, R. Morganti$^{3,4}$,
  B. H. C. Emonts$^{5}$ \\
$^{1}$Leiden Observatory, Leiden University, PO Box 9513, 2300 RA
  Leiden, The Netherlands.\\
$^{2}$Department of Physics and Astronomy, University of Sheffield,
  Sheffield, S3 7RH, UK.\\
$^{3}$Netherlands Institute for Radio Astronomy, Postbus 2,
7990 AA Dwingeloo, The Netherlands.\\
$^{4}$Kapteyn Astronomical Institute, University of Groningen, P.O. Box 800, 
9700 AV Groningen, The Netherlands\\
$^{5}$Australia Telescope National Facility, CSIRO Astronomy and Space
Science, PO
    Box 76, Epping NSW, 1710, Australia.}

\usepackage{ulem}

\newcommand{\kms}{km s$^{-1}$}

\defcitealias{villar99a}{Villar-Mart{\'{i}}n et al., 1999a}

\begin{document}
\maketitle
\begin{abstract}
We present new deep VLT/FORS optical spectra with intermediate
resolution and large wavelength coverage of the compact radio source
and ULIRG PKS 1345+12 (4C  12.50; z = 0.122), taken with the aim of
investigating the impact of the nuclear activity on the circumnuclear
interstellar medium (ISM). PKS 1345+12 is a
powerful quasar (L(H$\beta$)$_{\rmn{NLR}}$$\sim$10$^{42}$ erg s$^{-1}$)
and is also the best studied case of an emission line outflow in a
ULIRG.  Using the density sensitive transauroral 
emission lines  {[S II]}4068,4076 and  {[O
    II]}7318,7319,7330,7331, we pilot a new technique to
accurately model the electron density for cases in which it is not
possible to use the traditional diagnostic {[S II]}6716/6731, namely
sources with highly broadened complex emission line profiles and/or
high ($N_e \gtrsim$ 10$^4$ cm$^{-3}$) electron densities. We measure
electron densities of $N_e = (2.94_{-1.03}^{+0.71})\times10^3$
cm$^{-3}$, $N_e = (1.47_{-0.47}^{+0.60})\times10^4$ cm$^{-3}$ and 
$N_e = (3.16_{-1.01}^{+1.66})\times10^5$ cm$^{-3}$ for the
regions emitting the  narrow, broad and very broad components
respectively.  We therefore calculate a total mass outflow rate of
$\dot{M} = 8_{-3}^{+2}$ M$_{\sun}$ yr$^{-1}$, similar to the
range estimated for another compact radio source, PKS 1549-79
\citep{holt06}.  We estimate the total mass in the warm gas outflow is
$M_{\rmn{total}} = (8_{-3}^{+3})\times10^5$ M$_{\sun}$ with filling factors of
$\epsilon = (4.4_{-1.5}^{+1.8})\times10^{-4}$ and 
$\epsilon = (1.6_{-0.5}^{+0.7})\times10^{-7}$
 for
the regions emitting the broad and very broad components
respectively. The total kinetic power in the warm outflow is
$\dot{E}_{\rmn{total}} = (3.4_{-1.3}^{+1.5})\times10^{42}$ erg s$^{-1}$. Taking the
black hole properties published  by \citet{dasyra06}, we find that only a
small fraction ($\dot{E}/L_{\rmn{bol}} = (1.3\pm0.2)\times10^{-3}$) of the
available accretion power is driving the warm 
outflow in PKS 1345+12, which is significantly less than that
currently required by the majority of quasar feedback models
($\sim$5-10\% of $L_{\rmn{bol}}$), but
similar to the recent suggestion of \citet{hopkins10} if a two-stage feedback
model is implemented ($\sim$0.5\% of $L_{\rmn{bol}}$). The models also predict
that AGN-driven outflows will eventually remove the gas from the bulge
of the host galaxy. Our observations show that the visible warm
outflow in PKS 1345+12  is not currently capable of doing so. However,
it is entirely possible that much of the outflow is either obscured by
a dense and dusty natal cocoon and/or in cooler or hotter phases of
the ISM. This result is important not just for studies of young
(GPS/CSS) radio sources, but for AGN in general. 

\end{abstract}


\section{Introduction}
It has become increasingly clear that
the evolution and growth of supermassive black holes and their host
galaxy bulges are intricately linked, through the discovery of tight
correlations between the bulge and black hole properties 
(e.g. \citealt{ferrarese00,gebhardt00,tremaine02,marconi03,benson03}). 
However,
whilst
the observed correlations become better constrained, the physical
processes responsible are still not understood
(e.g. \citealt{cattaneo09}) with both  AGN feedback
(e.g. \citealt{silk98}) and  starburst supernova feedback
(e.g. \citealt{wild10}) providing plausible outflow driving mechanisms.

A number of theoretical models have successfully reproduced the
observed correlations by including AGN feedback
 in which  the
rapidly accreting  black hole dissipates energy into the
surrounding ISM by driving powerful outflows, which halt further
accretion and star formation  
(e.g. \citealt{silk98,fabian99,dimatteo05,springel05}). Currently, 
these  models require 
a significant amount of the accretion energy to be mechanically
coupled to the  ISM
($\sim$5-10\% of L$_{\rmn{bol}}$;
e.g. \citealt{dimatteo05,tremaine02,springel05,kurosawa09,booth09}),
although recent work by \citet{hopkins10} suggests that  if a two-stage feedback
model is implemented, the initial energy requirement may be a factor
of 10 lower.

Over the last decade, the signatures of such mechanical AGN feedback
(outflows) 
have been observed in Compact Steep Spectrum
(CSS; $D_{\rmn{radio}}$ $<$ 15kpc) and 
GigaHertz-Peaked Spectrum (GPS; $D_{\rmn{radio}}$ $<$ 1kpc) radio 
sources, both in the warm (e.g. optical emission line outflows;
e.g. \citealt{holt08}) and cold (e.g. HI absorption line outflows;
e.g. \citealt{morganti05}) gas. In the infra-red, emission line 
outflows have also recently been discovered in ULIRGs
\citep{spoon09}.

CSS and GPS radio sources 
are ideal objects for studying the effects of AGN feedback as they are 
 believed to be young, recently triggered radio-loud AGN (see
 \citealt{odea98} for an extensive review of these sources). The
 similarities in scale of the AGN, compact radio source and the dense 
 circumnuclear ISM mean that interactions will be strong and readily
 detectible. 
In 2006, Holt et al. performed a detailed study of
the outflow in the southern compact flat-spectrum radio source PKS
1549-79, deriving a mass outflow rate of  0.12 $< \dot{M} <$ 12
M$_{\sun}$ 
yr$^{-1}$ and a kinetic power of 5.1$\times$10$^{40}$ $< \dot{E} <
5.1\times10^{42}$ erg s$^{-1}$, which accounts for only a small
fraction of the available accretion energy in this source
(10$^{-6}$ $<$ $\dot{E}/L_{\rmn{Edd}}$ $\sim$ 10$^{-4}$). However, due
to the large uncertainties 
in the measured density, it was impossible to derive these crucial
parameters more accurately.

The young  radio galaxy PKS 1345+12 (GPS; $D_{\rmn{radio}}$ $\sim$350pc;
\citealt{stanghellini93}) is a prime object
for AGN feedback studies as it contains all of the signatures of a
 recently triggered 
powerful\footnote{ L(H$\beta$)$_{\rmn{NLR}}$$\sim$10$^{42}$ erg
   s$^{-1}$; c.f. the SDSS DR7 quasar catalogue \citep{schneider10} which
   includes H$\beta$ luminosities split into the NLR and BLR components. The
   NLR H$\beta$ luminosity of PKS
   1345+12 (we no not detect the BLR) is well above the mean value of
   NLR H$\beta$ luminosities in quasars, making it a very powerful
   AGN. This is also evident in its large {[O III]} luminosity. }
 AGN currently shedding its natal cocoon. Fast
 outflows   (up to 
$\sim$2000\kms)  are detected in both the warm and cold circumnuclear gas
 (e.g. \citealt{holt03}; H03 hereafter, \citealt{morganti05}). 
 The
 host galaxy is highly disturbed; the double nucleus (separated by
$\sim$1.8 arcsec or 4.3 kpc) is embedded in an extended, asymmetric
 halo with clear evidence for extended, distorted morphology
  (e.g. \citealt{smith89a}), suggesting PKS 1345+12 has been involved
  in a major merger in its recent past. In addition, prodigious star
  formation activity is observed. PKS 1345+12 has a 
  substantial far-IR excess ($L_{FIR}$ = 1.7$\times$10$^{12}$
  L$_{\odot}$, \citealt{evans99}), qualifying this source as an
  ultra-luminous IR galaxy (ULIRG). Furthermore, significant young 
 stellar populations have been detected in the halo, both in the
 diffuse emission 
 \citep{tadhunter05} and in a number of young Super Star Clusters (SSCs;
 \citealt{javi07}).  Hence, detailed studies of the outflow in PKS
 1345+12 are important for a number of classes of objects (powerful
 AGN, GPS/CSS radio galaxies and ULIRGs) with the outflow being the
 best studied example of an optical emission line outflow in a ULIRG.

A key parameter for determining the importance of the nuclear outflows
in terms of galaxy evolution is the electron density of the gas, which
allows the calculation of the mass outflow rate and kinetic power of
the outflow. 
Traditionally, gas densities have been successfully derived for many
AGN using the {[S 
    II]}6715,6731 doublet, which requires an accurate
measurement of the ratio between the two lines in the blend. 
However,  sources with highly complex emission line
profiles can preclude accurate modelling of the {[S  II]} doublet. In
the case of PKS 1345+12, the velocity widths and shifts of the broader
components are comparable to the separation of the doublet. Hence,
whilst total fluxes in each component can be established, it 
is often not possible to determine how much of the emission in each
component originates from each line in the blend.  In addition, our previous study of PKS
1345+12 (H03) has suggested that the densities in the nuclear regions of
young radio-loud AGN  are high ($N_{e}$ $>$ 4$\times$10$^{3}$
cm$^{-3}$), and 
may be high enough to be outside the range of densities to which the
[S II] diagnostic is sensitive ($N_e$ = 10$^{2}$-10$^{3}$ cm$^{-3}$).

In this paper, we present new, high quality VLT/FORS2 spectra of PKS
1345+12, which are both deeper, and  cover a wider wavelength range, 
than our previous WHT/ISIS
spectra presented in H03. The new spectra now include
  further density sensitive emission lines, namely the transauroral
  {[S II]}4068,4076 and {[O II]}7318,7319,7330,7331 lines,
to pilot a new technique  to measure the electron 
  density of the circumnuclear emission line gas.
   Consequently, we calculate for the first time the mass outflow rate and kinetic
  power of the optical (warm) outflow in a young radio-loud AGN and ULIRG. 

Throughout this paper we assume the following cosmology: H$_{0}$ = 71
\kms, $\Omega_{\rmn 0}$ = 0.27 and  $\Omega_{\Lambda}$ = 0.73.

\section{Observations and Data Reduction} 
\begin{figure*}
\centerline{\psfig{file=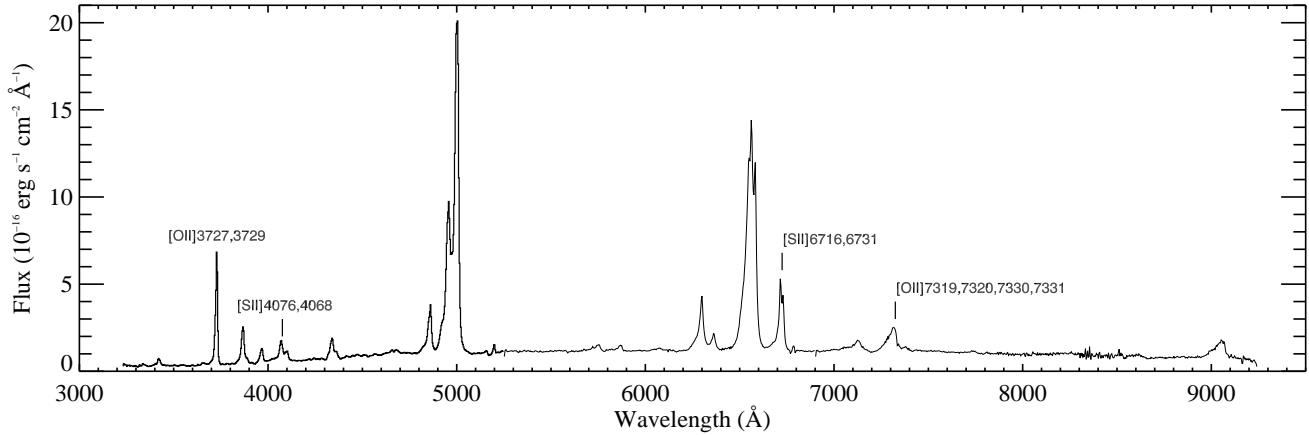,angle=0,width=19cm}}
\caption[]{Rest frame  spectrum of the nuclear aperture of
  PKS 1345+12. The key density diagnostic emission lines used in this
  paper are highlighted. 
}
\label{fig:spectrum}
\end{figure*}
We have obtained new, deep long-slit optical spectroscopic
observations of PKS 1345+12 in service mode on 22nd March 2007. Using
FORS2 on the ESO Very Large Telescope (VLT) in LSS mode with {\sc
  gris-600b+22}, {\sc gris-600ri+19} and {\sc gris-600z+23}, and a 1.3
arcsec slit, we
obtained spectra along PA 160 with  large spectral coverage (rest
frame: 3200-9200\AA).  

The data were reduced in the usual way (bias subtraction, flat
fielding, cosmic ray removal, wavelength calibration, flux
calibration) using the standard packages in {\sc iraf}. The
two-dimensional spectra were also corrected for spatial distortions of
the CCD. The final wavelength calibration accuracy, calculated uding
the standard error on the mean deviation of the night sky emission
line wavelengths from published values \citep{osterbrock96} was 0.281, 0.112
and 0.035 \AA~in the blue, red and {\it z} bands respectively. The
spectral resolution of the night sky emission lines was
6.65$\pm$0.16\AA, 7.15$\pm$0.41\AA~and 7.02$\pm$0.37\AA~in the blue,
red and z bands respectively.  

Comparison of several spectral photometric standard stars taken with a
wide slit (5 arcsec) gave a relative flux calibration accurate to
{$\pm$}5 per cent. This accuracy was confirmed by comparison to our
earlier data (H03). 
A further standard star was observed with a
narrow slit, matched to that used to observe PKS 1345+12, to correct
for atmospheric absorption features (e.g. A and B bands at $\sim$7600
and $\sim$6800\AA~respectively).

The main aperture used was the nuclear aperture -- 1.5 arcsec wide,
centred on the nuclear continuum emission, and is shown in Figure
\ref{fig:spectrum}. Before modelling the 
emission lines in the nuclear aperture, the continuum was modelled and
subtracted, a step which has been shown to be crucial in the modelling
of the broader, blueshifted components of the emission lines
(e.g. H03, \citealt{holt06,holt08,holt09}). Our continuum model
comprised a nebular continuum (see e.g. \citealt{dickson95}) and a
modelled stellar continuum following the methods
of e.g. \citet{tadhunter05,holt06,holt07}. 

The spectra were extracted and analysed using the {\sc starlink}
packages {\sc figaro} and {\sc dipso}.

\section{Results}
In this paper, our analysis focusses on the density of the emission
line gas in the circumnuclear regions. We will therefore discuss only
the relevant density diagnostic lines ({[O II]} and {[S II]}; see
Figure \ref{fig:spectrum}) and
refer the reader to 
H03 for a detailed discussion of the other strong nuclear
emission lines in this source.

\subsection{Emission line modelling}
The nuclear emission lines in PKS 1345+12 are
strong, characterised by highly broadened profiles with strong blue
wings. A detailed study of the emission line spectrum of the nucleus
of PKS 1345+12 showed that the emission lines required a minimum of 3
Gaussian components to model them (H03). Here, we implement the model
derived by H03 for {[O III]}4959,5007 (hereafter the
{[O III]} model).  This model comprises\footnote{The 
  notations narrow, intermediate, broad and very broad follow the
  kinematical component definitions of \citet{holt08}.}:
\begin{enumerate}
\item a narrow component, FWHM 340 $\pm$ 23 \kms~at the systemic
  velocity ($z$ = 0.12174 $\pm$ 0.00002);
\item a broad component, FWHM = 1255 $\pm$ 12 \kms;
  blueshifted by 402 $\pm$ 9 \kms~with respect to the narrow
  component (note H03 refer to this component as  `intermediate');
\item a very broad component, FWHM = 1944 $\pm$ 65 \kms, blueshifted by
  1980 $\pm$ 36 \kms~with respect to the narrow component (note H03
  refer to this component as `broad').
\end{enumerate} 
\begin{figure}
\centerline{\psfig{file=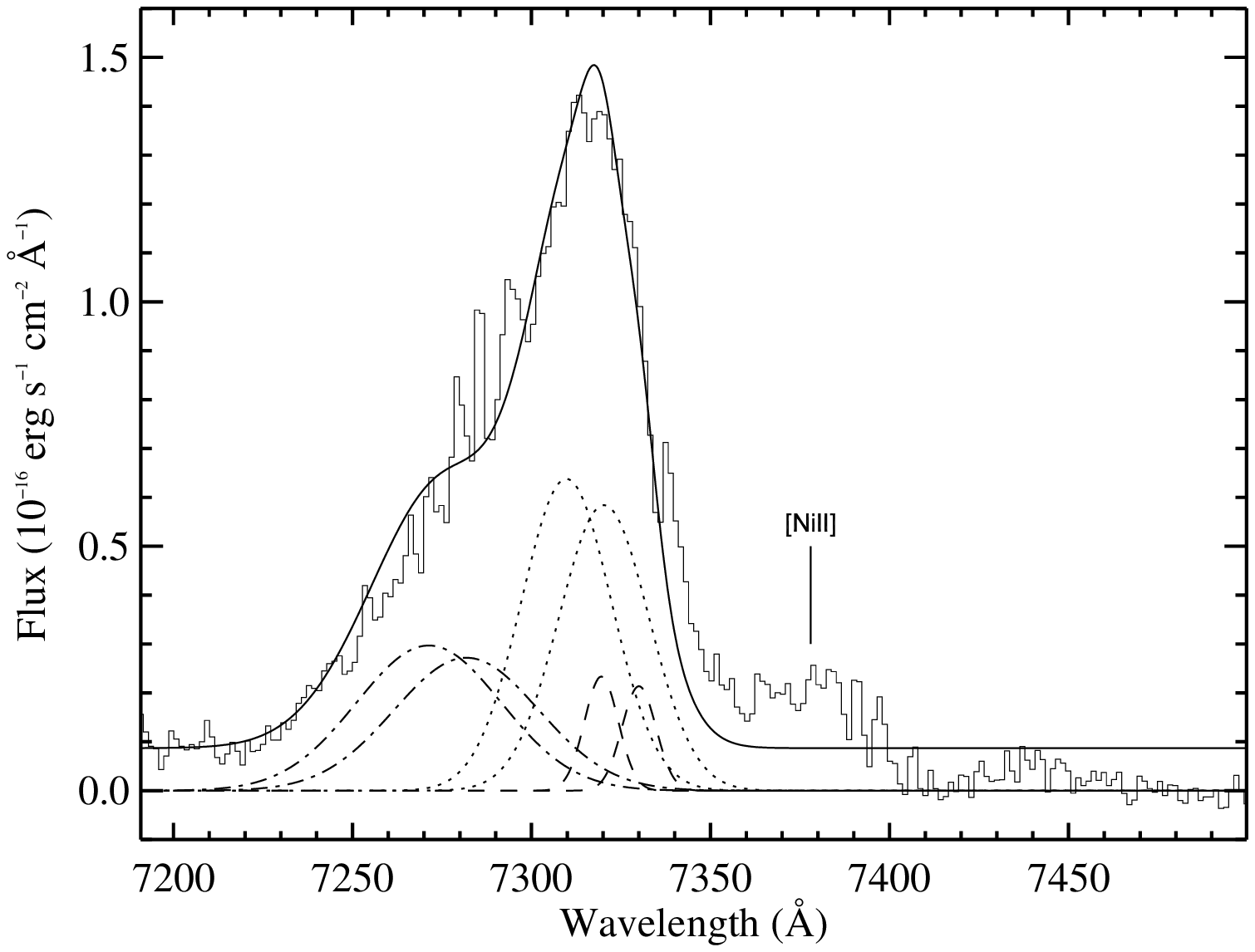,angle=0,width=9cm}}
\caption[]{
The {[O II]}7319,7320,7330,7331 emission line blend. The faint
solid line is the 
observed spectrum, the bold line is the overall fit, the dashed
lines are the narrow  components, the dotted lines are
the broad  components and the dot-dashed lines are
the very broad  components. The vertical line marks the
  position of {[Ni III]}7378 in the red wing of the {[O II]} blend.}
\label{fig:s2red}
\end{figure}
Figure \ref{fig:s2red} shows the {[O III]} model applied to the {[O
    II]}7319,7320,7330,7331 blend, which we have modelled as a doublet
(i.e. one line centred at $\sim$7320\AA~and one line centred at
$\sim$7330\AA), with each line comprising three distinct emission line
components. The {[O III]} model gives a good fit to
the blend. In addition, the {[O III]} model provides reasonable fits to all
of the other lines ({[S II]}6716,6731, {[O II]}3727,3729 and {[S
    II]}4068,4076). Note that {[O II]}3727,3729 and {[S II]}4076,4086 were also
modelled as single lines, for a number of detailed line-fitting reasons. 
 The measured line ratios are plotted on the density diagnostic 
diagram in Figure 
{\ref{fig:diagnostic}}.

\begin{figure*}
\centerline{\psfig{file=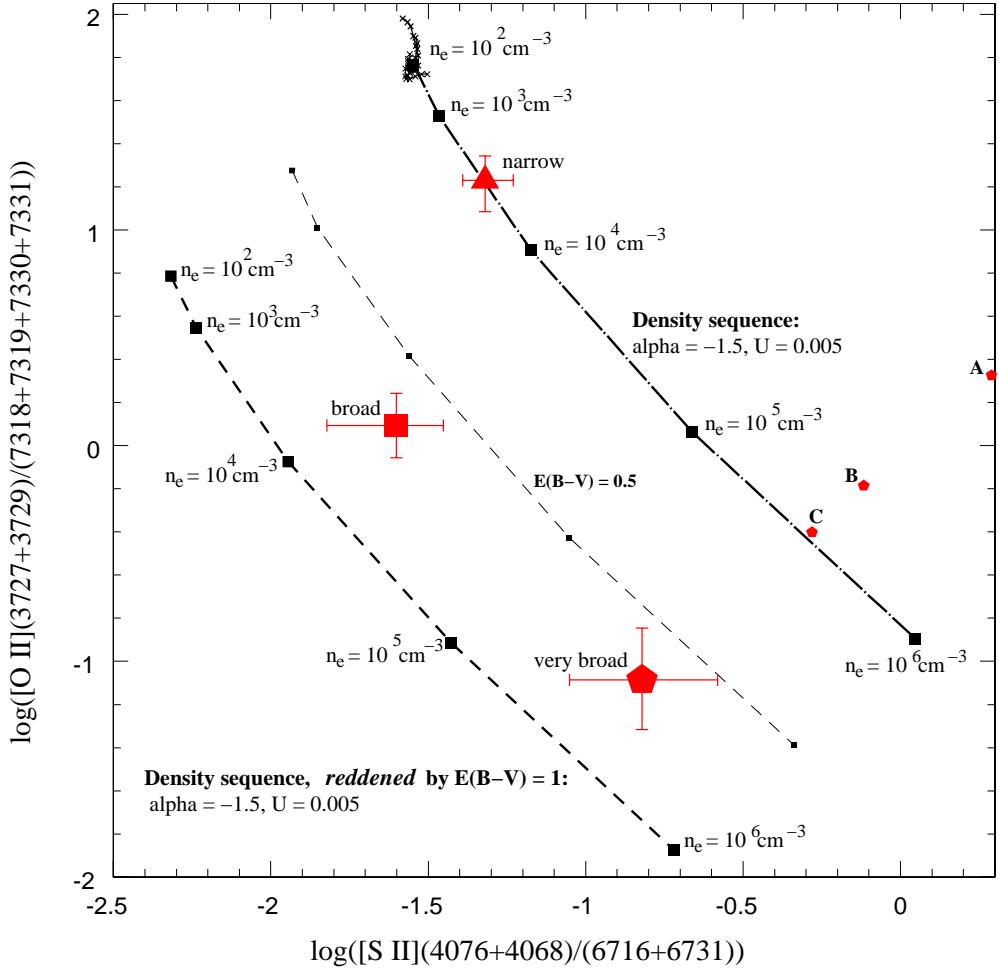,angle=0,width=14cm}}
\caption[]{ A diagnostic diagram using all of the key
density diagnostic lines. The crosses joined by solid lines are the
simple, optically thick AGN-photoionisation generated using the
multi-purpose photoionisation code {\sc mappings}. The models are
sequences in ionisation parameter, $U$, for
optically thick single-slab power-law (F$_{\nu} \propto \nu^{\alpha}$)
for spectral index  $\alpha$ = -1.0, -1.5, -2.0 and an electron
density of n$_e$ = 100 cm$^{-3}$. Also plotted is a density sequence
(squares joined by a 
dot-dashed line) which has a constant power law ($\alpha$ = -1.5) and
ionisation parameter ($U$ = 0.005) with electron densities of n$_e$ =
10$^2$, 10$^3$, 10$^4$, 10$^5$ and 10$^6$ cm$^{-3}$. Finally, the squares joined by the
dashed lines represent the same density sequence but {\it reddened} by
E(B-V) = 0.5 (faint line) and E(B-V) = 1.0 (bold line) using the
\protect\citet{seaton79} extinction law.  The
large red points are the measured (i.e. without reddening
correction) data: narrow (triangle), broad (square) and very broad
(pentagon). In addition, the small red pentagons give the position of
the very broad component after applying various reddening
corrections: A: E(B-V) = 1.44, B: E(B-V) = 0.92 and C: E(B-V) =
0.7. See the text for a discussion of the various E(B-V) values. }
\label{fig:diagnostic}
\end{figure*}

Note that H03 found that the {[O III]} model  did
    not model all of the emission lines perfectly. This  included {[S
      II]}6716,6731 which was best fit by a model with  slightly different FWHM
  and velocity shifts for the  broad and very broad
components. Furthermore, for all models to {[S II]}, the intensity
ratios for the components  needed to be fixed to ensure that they were
within the permitted range (0.44
$<$ {[S II]}6716/6731 $<$ 1.42) and that the fit was physically
viable.  The main
reason for the difficulty in modelling {[S II]} was that the widths and shifts of the
broader components are comparable to the separation of the
doublet. We find the  situation unchanged for the new VLT data. 
 For the new diagnostic, only the total flux in each kinematic component
is required. By removing the need for an accurate ratio between the
two lines in the [S II]6716,6731 doublet, the {[O III]} model can provide a good model
to the doublet. To test the robustness of our fit, in addition to our
`free intensity' {[O III]} model to the {[S II]}6716,6731 blend, we
have also tried forcing the component ratios so all lie within the
range quoted above. These give errors of 6\%, 9\% and 12\% on the
total fluxes in the narrow, broad and very broad components
respectively.

\subsection{The electron density}
A 
 key parameter for determining the importance of the nuclear outflows
in terms of galaxy evolution is the electron density of the gas. As
discussed above, due to the highly complex emission line profiles in
the nuclear regions of PKS 1345+12, it was not possible to measure the
density using the {[S 
    II]}6715,6731 doublet. Here, we pilot a new technique based
on the transauroral {[S II]} and {[O II]} lines.

Figure \ref{fig:diagnostic} shows a new diagnostic diagram constructed for
the transauroral density diagnostic lines {[O
    II]}3727/(7318+7319+7330+7331) and [S II](4068+4076)/(6716+6731).  As
discussed above, diagnostics relying only on the total flux in each
emission line component in each blend bypasses the issue faced with
the {[S II]}6716/6731 ratio. In addition, 
as demonstrated by the
reddening tracks plotted on  Figure
\ref{fig:diagnostic}, whilst the lines span a large range of
wavelengths, it is not necessary to perform separate reddening
corrections. The diagnostic is also not sensitive to the details of
the ionisation model. 

From Figure \ref{fig:diagnostic}, we obtain electron densities of
$N_e = (2.94_{-1.03}^{+0.71})\times10^3$ cm$^{-3}$, 
$N_e = (1.47_{-0.47}^{+0.60})\times10^4$ cm$^{-3}$ and
$N_e = (3.16_{-1.01}^{+1.66})\times10^5$ cm$^{-3}$ 
for
the narrow, broad and very broad components respectively. These
densities are much higher than the original estimates based on {[S
    II]}6716/6731 and suggest that, even without the problems faced
modelling the blend, the diagnostic would not give accurate densities
as the measured densities are beyond the range the {[S II]}6716/6731
ratio is sensitive to.

\begin{figure}
\centerline{\psfig{file=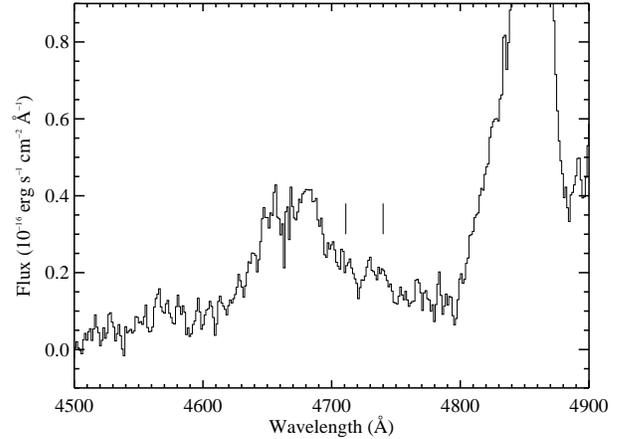,width=9cm,angle=0}}
\centerline{\psfig{file=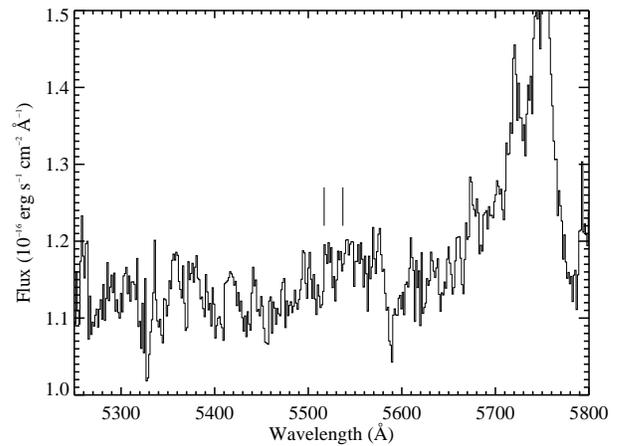,width=9cm,angle=0}}
\caption[]{The regions of the nuclear rest frame spectra in which the
  higher density sensitive lines {[Ar IV]}4713,4742\AA~(top panel)
  and{[Cl
      III]}5518,5534\AA~(bottom panel) would be located. Vertical
  lines mark the  wavelengths of the lines. A  weak
  component of {[Ar IV]}4742 is observed (top panel), but we do not
  detect any of the   other lines.  }
\label{fig:lines}
\end{figure}

Further diagnostic lines sensitive to higher densities are [Cl
  III]5518,5534 (sensitive to $\sim$ 10$^{4}$ cm$^{-3}$) and [Ar
  IV]4713,4742 (sensitive to 10$^4$ - 10$^5$ cm$^{-3}$). Both of
these doublets have been observed in Cygnus A, although they are weak
\citep{tadhunter94}. Neither doublet was detected in the original
      WHT/ISIS spectrum of H03. In these data, we detect very faint [Ar
  IV]4713,4742 although there is no evidence of
      [Cl III] emission (see Figure \ref{fig:lines}). The detection of
      {[Ar IV]} further supports the high densities measured in PKS
      1345+12. In addition, we detect  {[S III]}9069 (Figure 1) which
      also has a high critical density (1.2$\times10^6$
cm$^{-3}$).

\subsection{Reddening}

\begin{center}
\begin{table*}
\begin{minipage}{115mm}
\caption{Reddening values and emission line data. Columns are: $(a)$
  Quantity, where: H$\alpha$/H$\beta$ is the flux  
ratio between the measured values of H$\alpha$ and H$\beta$;
E(B-V)(H$\alpha$/H$\beta$) is the E(B-V) value    
 calculated using the H$\alpha$/H$\beta$ ratio and the standard
 interstellar extinction curve from 
\protect\citet{seaton79}; F$_{\rm H\beta}$ is the reddening corrected  H$\beta$ line flux 
(10$^{-15}$ erg s$^{-1}$ cm$^{-2}$); L$_{\rm H\beta}$ is the rest frame H$\beta$ luminosity 
(10$^{40}$ erg s$^{-1}$); 
E(B-V)(limits) gives the lower limits on the E(B-V) value derived for
the broad and very broad components by H03; E(B-V)(diagnostic)   is the E(B-V) value
derived from the diagnostic diagram in Figure \ref{fig:diagnostic};
$(b)$ value of
particular quantity for the narrow component; $(c)$ uncertainty for narrow component; column pairs
$(d)$ and $(e)$ and $(f)$ and $(g)$ are as for $(b)$ and
$(c)$ but for the broad and very broad components
respectively. \newline
$^a$ derived from considerations of the nebular continuum (H03). 
 $^{b}$ derived from H$\alpha$/H$\beta$ using the
model to the 
H$\alpha$,{[N II]} blend which gives the lowest possible flux in
H$\alpha$ (H03).  
$^c$
derived using the E(B-V) values given in row 2 of this Table. $^d$
luminosity from H03 updated to the cosmology used in this paper.   
$^e$ derived using the new (VLT) H$\beta$ fluxes and E(B-V)(diagnostic) values.}
\label{tab:reddening}
\begin{tabular}{lcccccc}\\ \hline 
& narrow & $\Delta$ &broad&$\Delta$ & \multicolumn{1}{c}{very broad} &$\Delta$\\
\multicolumn{1}{c}{$(a)$} & $(b)$ & $(c)$ & $(d)$ & $(e)$ &
\multicolumn{1}{c}{$(f)$} & $(g)$   \\ \hline \hline
{\bf Taken from H03} \\ 
H$\alpha$/H$\beta$          & 3.32 & 0.33 & 5.26 & 0.28 & 18.81 & 4.74\\
E(B-V)(H$\alpha$/H$\beta$)                   & 0.06 & 0.05 & 0.42 & 0.10 & 1.44  & 0.50\\
E(B-V)(limits) & & & $>$0.3$^a$ & & $>$0.92$^b$ \\
\\
F$_{\rm H\beta}$ (10$^{-15}$ erg s$^{-1}$ cm$^{-2}$) $^c$
&           0.72 & 0.06 & 8.8 & 0.4 & 43 & 10 \\
L$_{\rm H\beta}$ (10$^{40}$ erg s$^{-1}$)  $^d$
      &      2.7 & 0.2 & 33 & 2 & 160 &
40 \\
\\\hline
{\bf This paper} \\
E(B-V)(diagnostic)        & 0.0  & 0.1 & 0.7& 0.2 &0.7    & 0.2 \\
\\
F$_{\rm H\beta}$ (10$^{-15}$ erg s$^{-1}$ cm$^{-2}$) $^e$
& 1.1           & 0.2 & 41 & 5 & 15 & 4 \\
L$_{\rm H\beta}$ (10$^{40}$ erg s$^{-1}$)  $^{e}$
      &   4.3    & 0.7 &   155 & 18 & 56 &15 \\
\hline
\end{tabular}
\end{minipage}
\end{table*}
\end{center}

As discussed above, PKS 1345+12 is known to harbour a
dense and dusty circumnuclear cocoon. The level of reddening in the
optical spectrum was investigated in detail by H03, using three
separate techniques: Balmer line ratios (e.g. H$\alpha$/H$\beta$);
comparison of the optical results with  
published near-IR data (Pa$\alpha$/H$\beta$) and consideration of the
nebular continuum as part of the SED modelling. The reddening results
of H03 are summarised in Table \ref{tab:reddening}.  Here, we further
investigate the reddening using our density diagnostic (Figure \ref{fig:diagnostic}). 

In addition to electron density sequences, a number of reddening
(E(B-V)) sequences, derived using the \citet{seaton79} extinction law,
are also plotted on Figure \ref{fig:diagnostic}. 
From Figure \ref{fig:diagnostic}, the reddening is estimated to be 
E(B-V) = 0.0 $\pm$ 0.1, E(B-V) = 0.7 $\pm$ 0.2 and E(B-V) =  0.7 $\pm$
0.2 for the
narrow, broad and very broad components respectively. For the narrow
component, this is consistent with the results of H03 although the
E(B-V) values for the broad and very broad components are somewhat
higher and lower than those derived by H03 respectively. 

 In terms of line ratios, a `cleaner' method would be to use unblended
 line combinations, such as Pa$\alpha$/H$\beta$. H03 attempted to use
Pa$\alpha$/H$\beta$ to improve the reddening estimate but issues such
as differing components and uncertainties about possible slit losses
between the optical and near-IR data proved difficult.  We have since
obtained the unpublished near-IR spectrum of Veilleux (priv. comm.) in
an attempt to model Pa$\alpha$ with our three component {[O III]}
model. Whilst we can obtain a good fit, it is no better than the
original two component model, and still does not help with regards to
the issue of slit losses. 

Given the uncertainties regarding the degree of reddening, H03
attempted to derive limits 
for the reddening using a more indirect
technique.  By generating nebular continua with various E(B-V) values, H03 set a lower limit on
the reddening in the broad component, the component which dominates the
flux in the nebular continuum. This technique gave a lower limit 
of E(B-V) $>$ 0.3 for the broad component, which is consistent with
all reddening estimates for this component.

Given the uncertainties in the reddening estimates, we have also
plotted the reddening 
corrected line ratios for the very broad component using 
various E(B-V) values, in order to assess how the degree of reddening
affects our derived density. 
For this exercise we have chosen three E(B-V)
values:
\begin{enumerate}
\item point A: E(B-V) = 1.44; the value derived by H03 using
  H$\alpha$/H$\beta$;
\item point B:  E(B-V) = 0.92; the lower limit for the reddening
  derived by H03;
\item point C: E(B-V) = 0.7; the value derived by our diagnostic
  diagram.
\end{enumerate}
As can be seen from Figure \ref{fig:diagnostic}, whilst the poorly
constrained E(B-V) value will affect the reddening corrected H$\beta$
luminosity, it has little effect on the derived electron density. We
therefore conclude that the electron densities derived using this
technique are robust.

\section{Discussion}
\subsection{How important is the warm outflow in PKS 1345+12?}
In recent years AGN feedback, in the form of quasar-driven outflows, has been
relied upon to explain a number of observed relations, such as the
close correlation between black hole mass and the galaxy bulge
properties
(e.g. \citealt{silk98,fabian99,ferrarese00,gebhardt00,tremaine02,marconi03,dimatteo05,kurosawa09,booth09}) 
and the 
high end shape of the galaxy luminosity function
(e.g. \citealt{benson03}).   Currently, the majority of AGN feedback models require a
significant amount of the available accretion energy to power the
outflows ($\sim$5-10\%;
e.g. \citealt{dimatteo05,springel05,booth09,kurosawa09}) although this
has recently been brought into question, with much lower energy
injection required if a two-stage feedback model is implemented
($\sim$0.5\% of $L_{\rmn{bol}}$; \citealt{hopkins10}). 

PKS 1345+12 provides an ideal test  case for investigating the impact
of AGN feedback in the early stages of AGN evolution. Clearly
currently still undergoing a major merger/interaction (c.f. double
nucleus within a single diffuse halo with a number of tidal features
and evidence for recent star formation), the small, recently
triggered radio jets are currently expanding through a dense
natal cocoon of gas and dust (c.f. large electron densities and reddening, high
X-ray and HI absorbing columns). Our optical spectra reveal clear
signatures of AGN driven outflows in the strong, highly broadened and
blueshifted optical emission lines. Whilst the outflows are fast, it
is important to combine this kinematic information with other
properties, such as the gas density, to determine the key parameters
for understanding the importance of AGN feedback -- the mass outflow
rate and the kinetic power. 

Following the derivation in \citet{holt06}, the mass outflow rate for
a spherical outflow can be expressed as: 
\begin{equation}
\dot{M} = \frac{3 L(H\beta) m_p v_{out} }{N \alpha_{H\beta}^{eff} h \nu _{H\beta} r}
\end{equation}
where L(H$\beta$) is the H$\beta$ luminosity, $m_p$ is the mass of the
proton, $v_{out}$ the outflow velocity, $N$ is the electron density of
the gas, $\alpha_{H\beta}^{eff}$ is the effective H$\beta$ recombination
coefficient (see \citealt{osterbrock89}), $\nu_{H\beta}$ the frequency of
H$\beta$, $h$ the Planck constant and $r$ is the radius of the
spherical volume. 

We take 
H$\beta$ luminosities measured from our new VLT spectra, modelled with
the {[O III]} model of H03, and corrected for extinction using the
E(B-V) values derived by our diagnostic diagram (Figure
\ref{fig:diagnostic})  and the \citet{seaton79} extinction law, all of
which are  are summarised in Table
\ref{tab:reddening}.  
In our currently adopted cosmology, these 
are: L(H$\beta$)$_{\rmn N}$ = (4.3$\pm$0.7)$\times$10$^{40}$ erg
s$^{-1}$;
L(H$\beta$)$_{\rmn B}$ = (1.55$\pm$0.18)$\times$10$^{42}$ erg
s$^{-1}$ and 
L(H$\beta$)$_{\rmn {VB}}$ = (5.6$\pm$0.15)$\times$10$^{41}$ erg s$^{-1}$. 
 The rest frame outflow velocities ($v_{out}$) are $v_{\rmn B} =
$ 402$\pm$9 km s$^{-1}$ and  $v_{\rmn{VB}} =$ 1980$\pm$65 km s$^{-1}$.
Using our new diagnostic tool (Figure \ref{fig:diagnostic}), we
measure electron densities of $N_e = (2.94_{-1.03}^{+0.71})\times10^3$
cm$^{-3}$, $N_e = (1.47_{-0.47}^{+0.60})\times10^4$ cm$^{-3}$ and 
$N_e = (3.16_{-1.01}^{+1.66})\times10^5$ cm$^{-3}$
 for the narrow, broad and very
broad components 
respectively. 
If we assume case B recombination theory for an electron
temperature $T = 10,000$~K and that the outflow is spherical, with a
radius half of the diameter of the {[O III]}  emission line region  imaged by HST ACS
\citep{batcheldor07} (i.e. 7.5 milli-arcseconds or $r = 162$pc for
our cosmology),  we calculate mass outflow rates of 
$\dot{M} = 7.1_{-2.6}^{+2.0}$ M$_{\sun}$ yr$^{-1}$ and
$\dot{M} = 0.6_{-0.2}^{+0.3}$ M$_{\sun}$ yr$^{-1} $
in the two outflowing 
components respectively, with a total mass outflow rate
of $\dot{M} = 7.7_{-2.9}^{+2.2}$ M$_{\sun}$ yr$^{-1}$. 
 This estimate
is comparable to the range of mass outflow rates calculated for another
young radio source, PKS 1549-79 (H06) but lies between the mass
outflow rates estimated for the NLR of Seyferts (0.1-1.0 M$\odot$
yr$^{-1}$; e.g. \citealt{veilleux05}) and the much larger neutral
outflows detected in other ULIRGs (10$^{2}$-10$^{3}$
M$\odot$ yr$^{-1}$; e.g. \citealt{rupke05a,rupke05b}). 
 We estimate that the total mass in the warm gas
outflow is  $M_{\rmn{outflow}} = (8_{-2}^{+3})\times10^5$
M$_{\sun}$ with
filling factors\footnote{See equation 3 in \citet{holt08}. } of
$\epsilon = (4.4_{-1.5}^{+1.8})\times10^{-4}$ and
$\epsilon = (1.6_{-0.5}^{+0.7})\times10^{-7}$ 
 for the regions emitting the broad and very
broad components respectively. 

In order to gauge the impact of the warm gas outflow on the
circumnuclear gas, it is important to estimate the kinetic power of
the outflow,  including both the radial and turbulent components in the gas
motion. Assuming that the relatively large linewidth of the outflowing
gas reflects a turbulent motion that is present at all locations in
the outflow region, the kinetic power is: 
\begin{equation}
\dot{E} = 6.34\times10^{35}\frac{\dot{M}}{2}(v_{\rmn{out}}^2 +
({\rmn{FWHM}})^2/1.85) \;\; {\rmn{erg}}\: \rmn{s}^{-1} 
\label{eq:kinetic}
\end{equation}
where FWHM is the full width at half maximum of the 
[OIII] line component in km s$^{-1}$, 
and $\dot{M}$ is the mass outflow rate expressed in solar masses per
year. 
The kinetic powers are $\dot{E} = (2.3\pm0.9)\times10^{42}$~erg
s$^{-1}$ and 
$\dot{E} = (1.1_{-0.4}^{+0.6})\times10^{42}$~erg s$^{-1}$ for the broad and
very broad components. Hence, the
total kinetic power in the warm gas outflow is $\dot{E} 
= (3.4_{-1.3}^{+1.5})\times10^{42}$~erg s$^{-1}$.

Combining the kinetic power with the bolometric luminosity provides an
estimate of the fraction of the total available power from accretion
onto the black hole driving the outflows. \citet{dasyra06} publish two
black hole mass estimates for the active western nucleus of PKS
1345+12, one  derived using 
M$_{\rmn{BH}}$ -- $\sigma$ as M$_{\rmn{BH}}$ = 6.54$\times$10$^7$ M$_{\odot}$
\citep{dasyra06}, and a smaller mass estimate of  M$_{\rmn{BH}}$ = 1.75$\times$10$^7$
M$_{\odot}$ based on the assumption that $\frac{1}{2}$L$_{\rmn{FIR}}$ =
L$_{\rmn{bol}}$ = L$_{\rmn{Edd}}$\footnote{It is also possible to
  estimate the bolometric luminosity from the extinction-corrected {[O
      III]} emission line luminosity using the 
  conversion factors given in 
  \citet{lamastra09}. Our reddening corrected total {[O III]}
  luminosity is L$_{\rmn {[O III]}}$ = (2.11 $\pm$ 0.25) $\times$ 10$^{43}$ erg
    s$^{-1}$ and
    corresponds to a bolometric luminosity of L$_{\rmn {[O III]}}$ =
    (9.5 $\pm$ 1.1) $\times$ 10$^{45}$ erg s$^{-1}$, which is similar to the
    estimate derived from the FIR luminosity.}. Comparing the two
  black hole masses gives 
 an idea of the Eddington ratio: \begin{equation}
\frac{M_{\rmn{BH,FIR}}}{M_{\rmn{BH,dyn}}} =
    \frac{L_{\rmn{bol}}}{L_{\rmn{Edd}}} = 0.27
\end{equation}
This estimate for the Eddington ratio is reasonable. For the similar
source PKS 1549-79, \citet{holt06} derived Eddington 
ratios in the range 0.3 $<$ L$_{\rmn{bol}}$ $<$ 57 depending on the
black hole mass used. Furthermore, \citet{wu09} collated data from the
literature for a large sample of compact radio sources. Of the 51
sources with quoted L$_{\rmn{bol}}$/L$_{\rmn{Edd}}$\footnote{Whilst
  PKS 1345+12 is included in the paper, \citet{wu09} did not include
  the estimate of L$_{\rmn{bol}}$/L$_{\rmn{Edd}}$ from
  \citet{dasyra06}, which is expanded for clarity above.}, the
Eddington ratio is typically high (10$^{-2}$ $<$
L$_{\rmn{bol}}$/L$_{\rmn{Edd}}$ $<$ several) compared to samples of
radio loud and radio quiet quasars and broad line radio galaxies
(typically L$_{\rmn{bol}}$/L$_{\rmn{Edd}}$ $<$ 0.1; e.g. \citealt{dunlop03,marchesini04}) 
but similar to the ratios observed in narrow line seyfert
Is (L$_{\rmn{bol}}$/L$_{\rmn{Edd}}$ = 0.16;
\citealt{bian08}); with  L$_{\rmn{bol}}$/L$_{\rmn{Edd}}$ = 0.28 and
L$_{\rmn{bol}}$/L$_{\rmn{Edd}}$ = 0.26 for the CSS and GPS sources
respectively. The estimate of L$_{\rmn{bol}}$/L$_{\rmn{Edd}}$ = 0.27
for PKS 1345+12 derived from the ratio of the two black hole estimates
is therefore entirely consistent with other GPS sources where this
ratio is known with more certainty. 

With this in mind, we calculate the  fraction of
the accretion power 
in
  the outflow as:  $\dot{E}/L_{\rmn{bol}} = 
(1.3\pm0.2)\times10^{-3}$.  Hence, if the assumed bolometric
  luminosity conversion is correct,  only a small fraction of the available
accretion 
power is driving the warm gas outflow.  This is similar to the findings
for another compact radio source, PKS 1549-79 \citep{holt06}.   This is in stark contrast to
the requirements of the majority of quasar feedback models which require a much larger
fraction of the accretion power of the black hole ($\sim$5-10\% of
L$_{\rmn{bol}}$)  to be
thermally coupled to the circumnuclear gas, but similar to the recent
work of \citet{hopkins10}.

It is also unlikely that
the warm gas outflow  is 
capable of removing all the warm/cool gas from the central regions of
the host galaxy by itself. If we assume that PKS1345+12 has a total mass
$M_{\rmn{total}} 
\sim 10^{11}$~M$_{\odot}$, and  a gas mass $M_{\rmn{gas}} \sim
10^{10}$~M$_{\odot}$ contained within a radius of 5~kpc --
conservative assumptions for ULIRGs which are likely to have a larger
gas mass concentrated within a smaller radius -- the gravitational
binding energy of the gas is $E_{\rmn{bind}} \approx G M_{\rmn{gas}}
M_{\rmn{total}}/R_{\rmn{gas}} \approx 2\times10^{58}$~erg. In comparison,  the
warm gas outflow will deposit only $\sim 10^{56}$~erg
into the surrounding ISM, assuming that it can persist in its current
form for the typical $10^7$~yr lifetime of an extragalactic radio
source. 
Again, these observations are
consistent with the limits calculated 
 for another compact  (young) radio source PKS 1549-79 (H06).    This result 
may suggest that PKS1345+12 will grow into a large radio
 galaxy in a gas-rich host. However, with its double nucleus, PKS
 1345+12 is still in the pre-coalescence stage of the merger, observed
 during the first peri-centre passage, but immediately before (within
 $\sim$0.1 Gyr) of the coalescence of the nuclei
 \citep{tadhunter10}. According to the merger simulations, PKS 1345+12
 is therefore yet to pass through the 
 most active phase (both in terms of nuclear activity and star
 formation) of the merger when the nuclei coalesce. An increase in
 nuclear activity may increase the strength of the winds/outflows and
 an increase in star formation will also consume large quantities of
 the ambient ISM. It is therefore unclear whether or not PKS 1345+12 is a
 progenitor of a gas rich radio galaxy.

It is clear that the kinetic power of the warm gas outflow in PKS 1345+12
is significantly less than that required by the majority of current quasar feedback
models. Could this be because we have not accurately measured all of the
input parameters or made inaccurate assumptions? In the above
analysis, the key observable parameters 
are the electron density, the size of the outflow region, the
luminosity of the H$\beta$ line and the black hole parameters (black
hole mass and the Eddington ratio). Following through, an 
underestimate of the kinetic power, $\dot{E}$, would result from
an overestimated electron density and/or outflowing region size, or from an
underestimated H$\beta$ luminosity. Furthermore, if the black hole
mass and/or the bolometric luminosity is overestimated, the impact of the
outflow may be underestimated. 

We are confident that the size of
the outflowing region has not been overestimated as recent
high-resolution HST imaging of PKS 1345+12 have shown the optical line
emission to be on a similar scale to the compact radio emission
\citep{batcheldor07}. If anything, the outflowing region may be
significantly larger if projection effects are important. 

The
electron density derived by this new technique is  high ($N_e \sim
10^{4-5}$ cm$^{-3}$) and therefore one to two orders of magnitude
above typical NLR densities 
in the literature  
measured using traditional diagnostics. As discussed in Section 3.3,
the large uncertainties in the reddening estimate will  have a
significant effect on the line ratios. However, as demonstrated by
Figure \ref{fig:diagnostic}, even with
large changes in the reddening for the very broad component, the
derived density remains 
unchanged. Despite this, it is still important to consider whether
such high densities are realistic and what effect this has on our
overall conclusions.

Whilst it is possible the new
technique overestimates the density, very high densities are not
surprising; high gas densities have been observed in a number of
compact radio sources \citep{holt09} and the original study of PKS
1345+12  using the {[S
    II]}6716/6731 ratio calculated a relatively high lower limit ($N_e
\gtrsim 5\times10^3$ cm$^{-3}$; H03). Furthermore, large quantities of gas
in the nuclear regions of PKS 1345+12 have been observed in a variety
of wavebands (see Section 1).  Using the lower limit for the
electron density provides upper limits  for the mass outflow rate
($\dot{M} < 21$ M$_{\odot}$ yr$^{-1}$), the kinetic power ($\dot{E} <
3.8\times10^{43}$ erg s$^{-1}$) and the fraction of the available accretion power
in the outflow ($\dot{E}/L_{\rmn{Edd}} < 4\times10^{-3}$). Hence, even if
our electron density is overestimated by as much as two orders of
magnitude, the amount of energy in the outflow is still significantly
lower than the required 5-10\%. 

It is possible that the errors on the derived H$\beta$
luminosity are large.   As discussed
above, the nuclear regions of PKS 1345+12 are known to contain large
quantities of gas and dust. The optical emission line components are
known to be  heavily extinguished (H03 and Section 3.3), whilst radio and X-ray studies
detect large absorbing columns: $N_{\rmn{H}} > 10^{22}$ cm$^{-2}$
(radio; \citealt{morganti05}) and $N_{\rmn{H}} = 2.3\times10^{22}$
cm$^{-2}$ (X-ray; Siemiginowska priv. comm.). 
However,  as demonstrated in Section 3.3, it is difficult to constrain
the degree of reddening, although all estimates discussed in Section
3.3 do agree within the (large) errros. Above, we calculated the
various outflow properties, such as mass outflow rates and kinetic
powers, based on the H$\beta$ luminosities derived from the new
H$\beta$ flux measurements from the VLT spectra, and the reddening
estimates derived from the diagnostic diagram in Figure
\ref{fig:diagnostic}. We now calculate the same quantities using the
new H$\beta$ fluxes and the original reddening estimates from H03 in
order to understand the possible uncertainties in our results. 

In short, for this case, we calculate the following: (i) mass
outflow rates of $\dot{M}_{\rmn{B}}$ =
2.7 M$\odot$ yr$^{-1}$ and $\dot{M}_{\rmn{VB}}$ = 7.2 M$\odot$
yr$^{-1}$ with a total mass outflow rate
of $\dot{M}_{\rmn{total}}$ = 9.9 M$\odot$ yr$^{-1}$; (ii) volume
filling factors of $\epsilon_{\rmn{B}}$ = 1.7$\times$10$^{-4}$ and $\epsilon_{\rmn{VB}}$ = 
1.9$\times$10$^{-6}$; (iii) total gas mass of $M_{\rmn{total}}$ =
8$\times$10$^{5}$ M$\odot$; (iv) kinetic powers of 
$\dot{E}_{\rmn{B}} = 8.7\times10^{41}$~erg
s$^{-1}$ and 
$\dot{E}_{\rmn{VB}} = 1.4\times10^{43}$~erg s$^{-1}$ with a total
kinetic power of $\dot{E}_{\rmn{total}} =
1.5\times10^{43}$~erg s$^{-1}$; (v) fraction of the total available
accretion in the warm outflow: $\dot{E}$/L$_{\rmn{Edd}}$ =
6$\times$10$^{-3}$. Hence,  implementing the original reddening
estimate from H03 rather than those derived 
from Figure \ref{fig:diagnostic} decreases the impact of the broad
component by a factor of $\sim$3, but boosts the impact of the very
broad component by a factor of $\sim$12. Overall, this increases the
total impact of the outflow by a factor of 4.6, from 0.13\% of L$_{\rmn bol}$ to 0.6\% of
L$_{\rmn bol}$  and demonstrates that our results are 
robust against large variations in the reddening. 

The final possible source of uncertainties relates to the black hole
properties, namely the mass, the bolometric and Eddington luminosities
and the Eddington ratio . In the above
calculations, we have followed  \citet{dasyra06} and assumed that $\frac{1}{2}$L$_{\rmn{FIR}}$ = 
L$_{\rmn{bol}}$ = L$_{\rmn{Edd}}$ provides a good estimate of the
bolometric luminosity. Whilst this conversion may be a good general
assumption, in highly 
complex sources like PKS 1345+12, the FIR emission is likely to
contain many components. We have also  compared the black hole
masses in  \citet{dasyra06} and estimated an Eddington ratio of
0.27. To quantify the possible errors on our result, we  take the extreme case
and assume that PKS 1345+12 is accreting at the Eddington ratio, with
the larger of the two black hole masses. In this case, we calculate 
$\dot{E}/L_{\rmn{Edd}}$ = (3.6$\pm$0.16) $\times$ 10$^{-4}$. Hence,
potential errors in the black hole mass are likely to reduce the
impact of the outflow.

Given the above, as suggested for a similar source PKS 1549-79 by
\citet{holt06}, it is likely that  the warm gas 
outflow detected in the optical emission lines accounts for only a
small fraction of the total outflow with the remaining mass in hotter
or cooler phases of the ISM. Indeed \cite{morganti05} reported the
discovery of massive neutral outflows in a sample of 7 compact radio
sources which includes PKS 1345+12. 
The HI outflow in PKS 1345+12 is: $\dot{M} \sim 8-21$ M$_{\odot}$
yr$^{-1}$, with the lower end of the range  similar to that
for the warm gas outflow. Taking this range of mass outflow rates, and assuming the
outflow velocity and FWHM to be 600 \kms~(i.e. FWZI/2; Morganti
priv. comm.), we find the HI     outflow has a kinetic 
    power of $1\times10^{42} \lesssim \dot{E}$ (erg s$^{-1}$) $\lesssim
    4\times10^{42}$ which accounts for $4\times10^{-4} \lesssim
    \dot{E}/L_{\rmn{bol}} \lesssim 1\times10^{-3}$ 
of the available
    accretion energy.

Combining the HI and optical
    results, the total observed outflow is currently of the order of
    16-29 M$_{\sun}$ yr$^{-1}$, with a kinetic power of 4.4$\times10^{42}$ $<$
    $\dot{E}$ $<$ 7.4$\times10^{42}$ erg s$^{-1}$ and accounts for
    0.2-0.3\%   of the available
    accretion energy, L$_{\rmn{bol}}$.

It is also interesting to consider the amount of energy locked 
  into the radio emission by comparing the radio power,
  $Q_{\rmn{jet}}$, to the accretion energy. Taking the radio jet power
  calculated by \citet{wu09} of $log(Q_{\rmn{jet}}) = $
  44.62\footnote{At 178MHz.} and the
  above bolometric luminosity, we find $Q_{\rmn{jet}}/L_{\rmn{bol}}$ =
  0.16.  In other words, around 16 per 
  cent of the 
  total available accretion power is  
  currently powering the radio jets. Whilst there is various evidence
  to suggest that the HI and optical emission line outflows are driven
  by the radio jets (e.g. high outflow velocities e.g. \citealt{holt08}; co-spatial radio
  and optical emission on similar scales in GPS and CSS sources and in
  some CSS sources, evidence for radio-optical alignment
  e.g. \citealt{devries99,axon00,odea02,labiano05,batcheldor07,privon08}), it appears 
  that the mechanical coupling of the radio jet to the ISM is relatively 
  inefficient. 

PKS 1345+12 is also detected in  X-rays 
(e.g. \citealt{odea00,siemiginowska08}) with a soft X-ray excess
(L(0.5-2keV) = 2$\times$10$^{42}$ ergs sec$^{-1}$)
and a harder X-ray spectrum than typical for AGN  (Siemiginowska
priv. com.). 
The emission has two clear
features: i) an extended ($\sim$15 kpc) diffuse
component, which may be the result of thermal emission in the host
galaxy, and ii) an elongated feature to the SW of the active nucleus,
which appears to be closely related to the VLBI radio source. Further
observations are required to establish whether a hot gas outflow is
present in PKS 1345+12.

To summarise, after consideration of all possible sources of error,
the most extreme kinetic power we calculate is still at least an order
of magnitude smaller than that currently required by the majority of galaxy evolution
models but similar to the energy requirements suggested by recent work
on a two-phase feedback model \citep{hopkins10}.

\subsection{PKS 1345+12 -- a young radio source?}
It is now well accepted that compact (GPS and CSS) radio sources are
compact due to evolutionary stage rather than due to the effects of
frustration by a dense ISM (e.g. \citealt{fanti95}). However, recent
simulations of radio jet-ISM interactions have shown that, if the ISM forms
 dense clouds, much smaller masses could significantly distrupt
the propagation of the jet. For example, 
jet interactions with dense clouds  (e.g. of the order $M_{gas} \sim 10^{6}$
M$_{\odot}$) may be significantly disrupted (e.g. \citealt{sutherland07})
compared to the much larger gas masses required ($M_{gas} \sim 10^{9-11}$
  M$_{\odot}$) required if the gas is distributed 
in a homogeneous or clumpy distribution
(e.g. \citealt{deyoung93,carvalho94,carvalho98}). 
 
We have seen that the nuclear regions of PKS 1345+12 harbour a rich,
dense ISM, but
is there sufficient mass to confine and frustrate the radio source?
 
In our original study in 2003, using the following equation: 
\begin{equation}
\rmn{M_{gas}} = \rmn{m_p} \frac{\rmn{L}(\rmn{H}\beta)}{\rmn{N_e}
\alpha_{\rmn{H}\beta}^{eff} h\nu_{\rmn{H}\beta}}
\end{equation}
we derived  limits on the gas mass for PKS 1345+12 of:
M$_{\rmn{gas}}$ $>$ 2.61$\times$10$^{5}$ M$_{\odot}$, M$_{\rmn{gas}}$
$<$ 0.92$\times$10$^{5}$ 
M$_{\odot}$ and  M$_{\rmn{gas}}$ $<$ 5.64$\times$10$^{5}$ M$_{\odot}$
in the regions 
emitting the narrow, broad and very broad components respectively;
with an upper limit of order 10$^{6}$
M$_{\odot}$ for the total mass of line emitting gas in the
kinematically disturbed emission line components.

Using the new gas densities and H$\beta$ luminosities measured in this
paper, we now derive gas  
masses of: M$_{\rmn{gas}}$ = 
(1.2$_{-0.4}^{+0.3}$)$\times$10$^{5}$ M$_{\odot}$, M$_{\rmn{gas}}$ =
(7.7$_{-2.6}^{+3.1}$)$\times$10$^{5}$ M$_{\odot}$ and M$_{\rmn{gas}}$ =
(2.9$_{-0.6}^{+1.0}$)$\times$10$^{4}$ M$_{\odot}$ for the narrow, broad and very broad
components respectively. This gives a total gas mass of
(9.2$_{-3.0}^{+3.4}$)$\times$10$^{5}$ M$_{\odot}$ with
(8$\pm$3)$\times$10$^{5}$ M$_{\odot}$ 
in the outflowing components.

H03 discussed that, even with the original estimate of $<$10$^{6}$
M$_{\odot}$, it was unlikely that there was sufficient {\it warm}
gas in the nuclear regions of PKS 
1345+12 to confine and frustrate the young, expanding radio jets and
only the presence of dense clouds would be able to significantly disrupt the propagation of
the jet. The newly derived gas densities suggest the {\it warm} gas mass is one-two
orders of magnitude smaller than the previously calculated limit. It
is therefore unlikely that the propagation of the radio jets in PKS 1345+12 will be
significantly affected by the {\it warm} ISM in this source. However,
it should be noted that much higher gas masses have been measured in
e.g. molecular gas (3.3$\times10^{10}$ M$\odot$; \citealt{evans99}).

\section{Conclusions and Future Work}
Using new, deep VLT/FORS spectra of the young radio source PKS
1345+12, we have piloted a new technique using the transauroral {[S
    II]} and {[O II]} emission lines to determine the electron
density in a source where the highly broadened and complex emission
line ratios precluded measurements using the traditional
diagnostics (e.g. {[S II]}6716/6731). Our observations reveal that the
outflowing warm gas in 
this compact radio source is dense ($N_e \sim 10^{4-5}$ cm$^{-3}$). In
addition to previous issues regarding the modelling of {[S
    II]}6716,6731 doublet, this shows that the electron densities in
this source are much higher than the range of densities to which the
diagnostic is sensitive. 

The newly determined electron densities have
enabled us to calculate key parameters for determining the impact of
the fast, warm outflow in this source (mass outflow rate, kinetic
power). Whilst at face value the warm gas outflow appears extreme,
with outflow velocities of up to 2000 \kms, the outflow is driven by
only a small fraction of the energy available from accretion
power. Comparisons with the majority of AGN feedback models in the
literature suggest that the outflow in PKS 1345+12 is at least an
order of magnitude smaller than required, However,  recent work by
\citet{hopkins10} suggests that with a two-stage feedback model, the
initial feedback requirements may be similar to what is observed in
PKS 1345+12.

\section*{\sc Acknowledgements}
Based on observations collected at the European Southern Observatory, Chile
(078.B-0537). JH acknowledges financial support from NWO.  We thank
Prof. S. Veilleux for making the near-IR spectrum available to us. We
thank Dr. C. Booth for useful discussions regarding AGN feedback theory.

\bibliographystyle{mn2e}
\bibliography{abbrev,refs}

\appendix

\end{document}